\documentclass[prl,aps,reprint,amssymb,superscriptaddress]{revtex4-1}
\usepackage{graphicx}
\usepackage{amsmath}
\usepackage{dcolumn}
\usepackage{bm}
\usepackage{amsfonts}
\usepackage{amssymb}
\usepackage{CJKutf8}
\usepackage[thinspace,squaren,pstricks,derivedinbase,derived]{SIunits}

\begin{document}
\title{Proposal for electron quantum spin Talbot effect}

\author{W. X. Tang}
\email[Corresponding author: ]{wenxin.tang@monash.edu}
\author{D. M. Paganin}
\affiliation{School of Physics, Monash University, Victoria 3800,
Australia }

\begin{abstract}

We propose a quantum spin Talbot effect for an electron beam
transmitted through a grating of magnetic nanostructures. Tunable
periodic magnetic nanostructures can be used in conjunction with
electron-beam illumination to create a spin polarized replica of
the transversely periodic exit surface beam a Talbot length away,
due to quantum interference. Experiments have been proposed to
verify the effect in a two dimensional electron gas. This effect
provides a new route to modulate electron spin distributions
without a magnetic field. A quantum spin Talbot interferometer and
transistor are proposed for spintronics applications.

\end{abstract}
\maketitle

The ability to tune scalable semiconductor-based spintronics
devices, based on the intrinsic spin of electrons to store and
manipulate information, is both important and highly challenging
for spin-based electronics since spin injection, spin accumulation
and spin modulation of electrons are required
\cite{peter,albert,prinz1,prinz2,wolf}. Currently, manipulation of
the spin during transport between injector and detector via spin
precession and spin pumping can be accomplished \cite{spin},
however, those methods have difficulty controlling spin
distributions. By contrast, local tunability of spin distributions
over nanometer scales is crucial for future solid state quantum
computers based on electron spin \cite{andrea}. Inspired by the
progress in fabricating and controlling nanoscale magnetic
structures \cite{parkin}, we propose a spin-dependent quantum
Talbot effect for electron waves transmitted by a grating composed
of magnetic nanostructures, to modulate the spin lattice pattern
formed from a spin polarized replica of the structure upon
propagation through a Talbot length period and adjustable by
controlling the electron wavelength and magnetic nanostructures'
period. This leads to potential applications such as a quantum
spin Talbot transistor and a quantum spin Talbot interferometer.

The optical Talbot effect was discovered in 1836 \cite{Talbot},
and later explained by Rayleigh as a natural consequence of
Fresnel diffraction. He showed that the Talbot length
$\mathcal{Z}$$_{T}$ is given by
$\mathcal{Z}_{T}$=$\frac{2a^{2}}{\lambda}$ \cite{Rayleigh}, in the
paraxial approximation $a\gg\lambda$, where $a$ is period of the
grating and $\lambda$ is the wavelength of the incident light.
However, in a non-paraxial regime where
$\lambda$$\leqslant$$a$$<$2$\lambda$, the Talbot effect is also
operative for nonevanescent components of the scattered beam
\cite{eero}. This effect reveals the wave-nature of both radiation
and matter wave fields, examples of the latter including atoms,
electrons and plasmons \cite
{chapman,cronin,benjamin,eero,dennis,leskova}.

\begin{figure}[h!]
\includegraphics[width=0.4\textwidth]{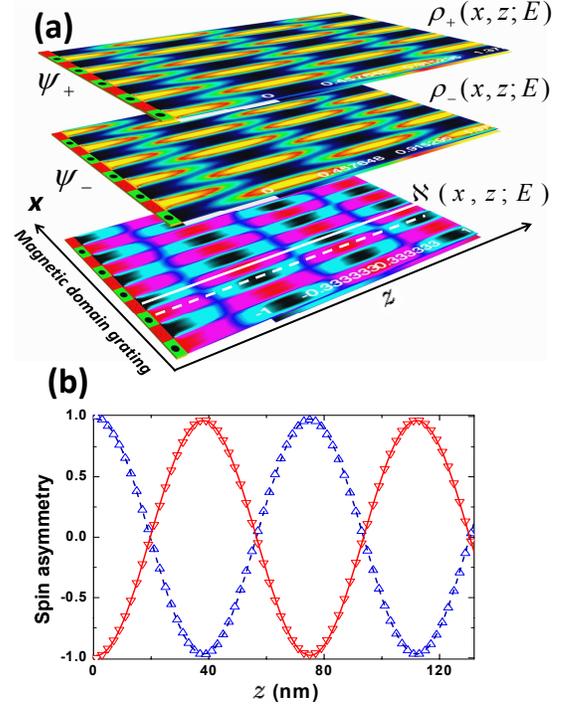}
\caption{Normalized diffraction intensity and profiles. (a) Maps
for spin up $\rho_{+}$ and down $\rho_{-}$ probability density
with spin asymmetry $\aleph(x,z; E)$. (b) Two spin asymmetry
profiles along $z$ are indicated by symbols $\bigtriangleup$ and
$\bigtriangledown$, and fit by Eq. \ref{Talbot_fit}, where
 $a$ = 20 nm, $\lambda$ = 10 nm, $\mathcal{Z}$$_{T}$ = 75 nm from Eq. \ref{Talbot_distance}.}
 \label{rys3}
\end{figure}

In this Letter, we calculate a spin polarized non-paraxial Talbot
effect for electron matter waves transmitted through a grating
composed of magnetic nanostructures. We find that the spin
asymmetry of the transmitted field varies with distance from the
grating, creating an electron spin replica of the structure a
Talbot length away, in a non-paraxial regime where
$\lambda$$\leqslant$$a$$<$2$\lambda$. This creates a tunable spin
lattice in two-dimensional space, which is a powerful method to
manipulate electron spin distributions in solid state systems. We
find that the quantum spin  interference pattern strongly depends
on the wavelength and grating period. Based on our theoretical
results, we propose experiments to verify this quantum spin Talbot
effect (QSTE) in a two dimensional electron gas (2DEG) system and
an atomically flat surface by spin polarized scanning probe
microscopy. We propose both a quantum spin Talbot interferometer
(QSTI) and quantum spin Talbot transistor (QSTT) devices.

For a grating with period $a$, normally illuminated with a
monoenergetic electron plane wave, the two-component spatial
electron wave function $\{\psi_+,\psi_-\}^T$ at energy $E$ and at
any distance $z \ge 0$ downstream of the exit surface $z=0$ is
\begin{eqnarray}\label{wavefunction}
    \psi_{\pm}(x,z;E) = \sum_m c_m^{\pm}(E) \exp\left[ \dot{\imath}(\gamma_{m}x+t_{m}z)\right].
\end{eqnarray}
\noindent Here, $x$ is the transverse coordinate, $c_m^{\pm}(E)$
denotes the Fourier coefficients of the two independent electron
spin projections, $\gamma_{m}$=$\frac{2\pi m}{a}$, and
$t$$_{m}$=$\sqrt{(\frac{2\pi}{\lambda})^{2}-\gamma_{m}^{2}}$
\cite{eero,dennis,leskova}, $\lambda=h/\sqrt{2m_eE}$ is the de
Broglie wavelength, $h$ is Planck's constant and $m_e$ is the
electron mass. $+$ and $-$ represents ``spin up''and ``spin down''
states of electron spin, respectively.

Consider a grating formed by nanoscale magnetic structures, for
example, magnetic stripe domains, as shown in Fig. 1. Electron
waves have a different complex transmission coefficient depending
on the configuration of the incoming electron beam spin state
relative to the magnetization direction of magnetic domains
(parallel ($\uparrow\uparrow$, $\downarrow\downarrow$) or
anti-parallel ($\uparrow\downarrow$, $\downarrow\uparrow$))
\cite{peter,albert,wolf}. Consequently, the spin up (down)
electron wave $\psi_{+}$($\psi_{-}$) propagates through the up
(down) magnetic domains in the grating, therefore achieving
separation of the electron wave depending on the spin state is
expected as shown in Fig. 1(a). The electron wave $\psi_+$ passes
through the magnetic ``up'' domains $A$ (green) with 100\%
transmission while being blocked completely by domains $B$ (red).
The color denotes the magnetization direction of a single domain.
The corresponding probability density diffracted from the magnetic
grating is longitudinally periodic in $z$ with period
$\mathcal{Z}_{T}$. Simultaneously, the probability density depends
on the spin of the electron wave shown in Fig. 1(a), as given by:
\begin{eqnarray}\label{probability_density}
   \rho_{\pm}(x,z; E) = \sum_m \sum_n c_m^{\pm *}(E) c_n^{\pm}(E)
   H_{m,n}(x,z;E),
\end{eqnarray}
\noindent where
\begin{eqnarray}\label{Hmn}
   H_{m,n}(x,z;E) = \exp\{\dot{\imath}[(\gamma_{n}-\gamma_{m})x+(t_{n}-t^{\ast}_{m})z]\}.
\end{eqnarray}
\noindent By definition, the spin asymmetry $\aleph(x,z;E)$
$\equiv$ ($\rho_+$ -$\rho_-$)/($\rho_+$ +$\rho_-$) will have the
same longitudinal periodicity as the probability density. Since
both the numerator and the denominator have a longitudinal
periodicity equal to the Talbot distance, when either
$a$$\gg$$\lambda$ or $\lambda$$\leqslant$$a$$<$2$\lambda$,
$\aleph(x,z;E)$ implies a {\em continuously tunable spin lattice}
in two dimensional space as shown in Fig. 1(b). The distribution
of two-dimensional spin asymmetry is determined by $\lambda$ and
$a$. In Fig. 1, $\lambda$=10 nm and $a$=20 nm; the numerically
calculated Talbot distance $\mathcal{Z}$$_{T}$ is 75 nm instead of
80 nm as expected by the conventional formula
2$a$$^{2}$/$\lambda$. The discrepancy is due to the paraxial
approximation in conventional Talbot theory. To calculate the
non-paraxial $\mathcal{Z}$$_{T}$ by the self-imaging condition
$\aleph$($x$,$n$$\mathcal{Z}$$_{T}$;$E$)=$\aleph$($x$,0;$E$), for
integer $n$ and $\lambda$$\leqslant$$a$$<$2$\lambda$, we obtain

\begin{figure}[h!]
\includegraphics[width=0.4\textwidth]{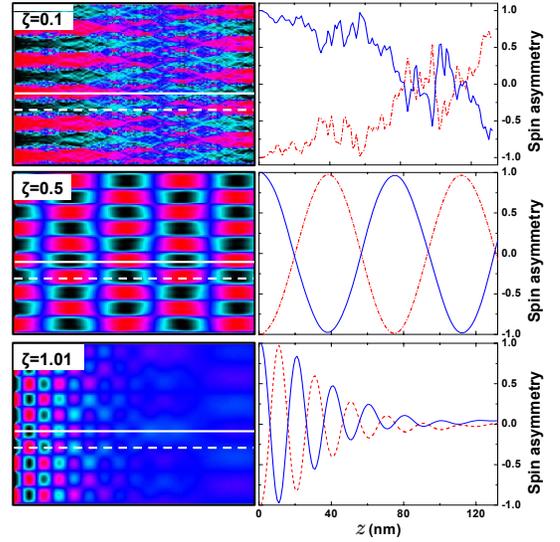}
\caption{Two dimensional spin-distribution $\aleph$($x$,$z$;$E$)
and corresponding profiles for $\zeta$ = 0.1, 0.5 and 1.01,
respectively.} \label{rys3}
\end{figure}

\cite{eero},
\begin{eqnarray}\label{Talbot_distance}
\mathcal{Z}_{T}=\frac{\lambda}{1-[1-(\lambda/a)^{2}]^{1/2}}.
\end{eqnarray}
\noindent From Eq. \ref{Talbot_distance}, $\mathcal{Z}$$_{T}$=
74.6 nm, consistent with our numerical results based on Eqs. 2--4.
For the paraxial approximation $a$$\gg$$\lambda$, Eq.
\ref{Talbot_distance} approaches 2$a$$^{2}$/$\lambda$, consistent
with the literature \cite{eero,dennis,leskova}. The spin asymmetry
distribution (Fig. 1 (b)) in the range 0.5$<$$\zeta$$\leqslant1$,
where $\zeta$=$\lambda$/$a$, is
\begin{eqnarray}\label{Talbot_fit}
\aleph(x,z;E)=A_{0}(x)\sin\frac{2\pi z}{\mathcal{Z}_{T}},
\end{eqnarray}
\noindent where $A_{0}$(x)=$\sin$$\frac{2\pi
x}{a}$$\cdot$$[$$\frac{\pi}{8}$+$\frac{2}{\pi}$$\sin$$^{2}$($\frac{2\pi
x}{a}$)$]$$^{-1}$. Note that evanescent waves have been neglected
in calculating the above expression. If $x$= $a$/4, then
$A_{0}$$\approx$0.97 as shown in Fig. 1(b). Therefore, spin
lattices can be tailored through nm to sub-$\upmu$m depending on
$\zeta$ and $a$.

To understand this tunability, we calculated the QSTE for
different $\lambda$. We find that the spin asymmetry profile
curves show a simple sine relationship when 0.5$<\zeta\leqslant1$.
However, if $\zeta<0.5$, the curves have complex structures and
small ripples decorate the spin asymmetry distribution (e.g.
$\zeta$= 0.1); when $\zeta$ $>$1, evanescent waves imply that the
polarization of spin decreases exponentially along $z$ (e.g.
$\zeta$= 1.01); Eqs. \ref{Talbot_distance} and \ref{Talbot_fit}
are not applicable for these ranges. Movie 1 shows sequential
evolution of the $\rho_{\pm}$ and $\aleph$  with $\zeta$ at $a$=
20 nm \cite{movie1}.

In a more realistic model, the electron wave undergoes partial
transmission at the antiparallel configuration between spin
orientation and magnetization direction of the domain. Considering
this, we find that the probability density distributions are
blurred but nevertheless distinguishable; even assuming only 1\%
transmission difference (TD) between two channels. The intensity
contrast and spin polarization drop with TD by the same order of
magnitude (Fig. 3).

\begin{figure}[h!]
\includegraphics[width=0.4\textwidth]{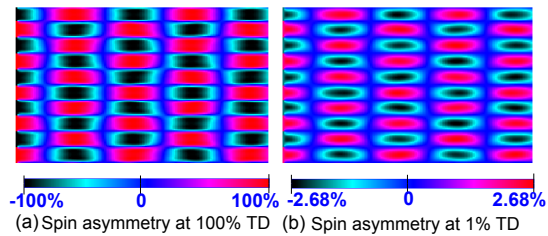}
\caption{The spin asymmetry distribution dependent on
transmission-rate difference between two channels. $a$=20 nm and
$\lambda$=10 nm.} \label{rys3}
\end{figure}

For a finite energy spread, assume an incident distribution of
electron energies $S_i(E)$. Under this model, we obtain
$\overline{\aleph}$($x$,$z$)$\equiv$$\int
S_i(E)$$\aleph$($x$,$z$;$E$)$dE$/$\int S_i(E)$$dE$. The influence
of energy spread on the 2D QSTE is calculated by numerical
evaluation of $\overline{\aleph}$($x$,$z$), assuming $S_i(E)$ to
be uniform from $\lambda$=15 nm to 20 nm, with the results shown
in Fig. 4. Surprisingly, a dramatic longitudinal modulation of
spin polarization near the grating is observed. Consequently, the
spin Talbot distance is also modulated depending on the energy
spread $S_i(E)$. In Fig. 4, instead of one peak appearing within
each spin Talbot distance, multiple peaks appear. Therefore, to
verify the QSTE, a narrow energy spread is desired, or the method
should have high energy resolution to distinguish different energy
channels.

A 2DEG at interfaces such as in a AlGaAs/GaAs heterostructure is a
candidate for testing the effect due to the small energy spread at
the Fermi level. In addition, the high mobility of electrons ($>$
3x10$^{6}$ cm$^{2}$V$^{-1}$s$^{-1}$) and their long spin
transportation distance ($>$ hundred $\upmu$m) are suitable
properties for spatial imaging of this effect
\cite{holland,DD,wolf,TP,mark}. The electron de Broglie wavelength
at the Fermi energy is unusually long, around 20--100 nm
\cite{TP}, making it easy to design a suitable magnetic domain
period $a$ and minimize effects caused by nonzero domain wall
width \cite{schmid,wxtang}. Further concern includes suitable
materials for the grating formed by magnetic stripe domains. The
wavelength of electrons in metal is normally less than 1 nm,
therefore, dilute magnetic semiconductors such as MnGaAs might be
suitable to form magnetic domain gratings with similar band
structure to AlGaAs/GaAs. Furthermore, by applying a pulsed
electron current along the grating, the period $a$ is tunable by
domain wall motion in ns \cite{shinjo,parkin}.

\begin{figure}[h!]
\includegraphics[width=0.4\textwidth]{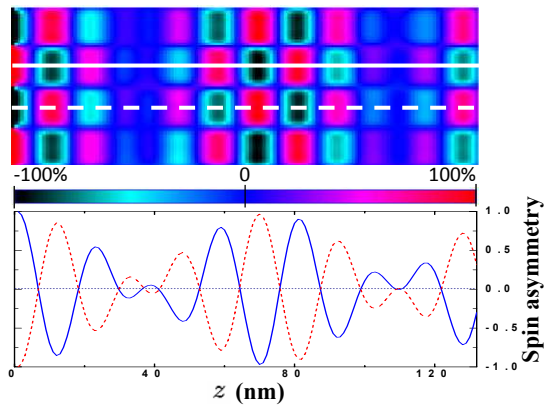}
\caption{Polyenergetic spin asymmetry corresponding to
$S_{i}(E)$=constant in range $\lambda$= 15--20 nm, using 800
integral steps in the numerical integration; $a$= 20 nm.}
\label{rys3}
\end{figure}

We have described the properties of a 2DEG system for testing the
effect, however, a spin dependent spatial imaging method is
needed. A spatial imaging technique has been elegantly applied in
spin Hall-effect detection in a 2DEG system by Scanning
Magneto-optic Kerr microscopy (SMOKE) \cite{spinhallmoke}. This is
an ideal way to demonstrate the QSTE.

Besides SMOKE, imaging electron flow in a 2DEG at the nanoscale
has been achieved based on a scanning probe method
\cite{mark,spm}. This measure can be applied to verify the QSTE at
GaAs/AlGaAs interfaces. In particular, scanning tunnelling
microscopy with a spin polarized tip (SP-STM) is an ideal
technique for investigating the surface electron wave QSTE
\cite{binnig,wiesendanger,oka}, as it provides both spin contrast
and atomic resolution. Recently, spin-dependent quantum
interference within a Co magnetic nanostructure by SP-STM has been
reported \cite{oka}. Inspired by this experiment, we believe
SP-STM could be used to see QSTE on an atomically flat surface by
a grating formed by an antiferromagnetic atomic chain. One of the
advantages of SP-STM is its high energy resolution in $dI$/$dV$
spin-asymmetry spectra to differentiate energy channels
\cite{oka}.

\begin{figure}[h!]
\includegraphics[width=0.45\textwidth]{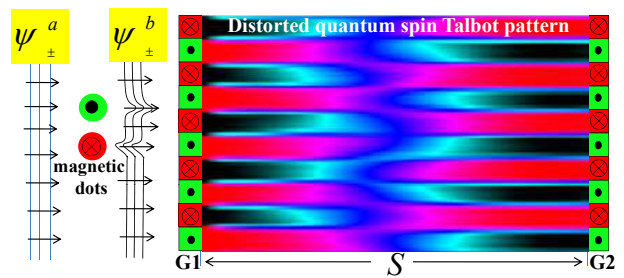}
\caption{Schematic representing deformation of the quantum spin
Talbot pattern caused by the spin-dependent scattering at magnetic
dots in front of a grating.} \label{rys3}
\end{figure}

We note that a far field non-spin electron Talbot interferometer
has been achieved \cite{benjamin}. Given the quantum spin Talbot
pattern obtained in our work, it is straightforward to formulate a
quantum spin Talbot interferometer (QSTI)(Fig. 5). In contrast to
the non-spin Talbot case, the QSTI is very sensitive to the change
in magnetization of magnetic dots located at the front of the
grating. This is extremely valuable to read out the magnetic
configuration of such dots. The QSTI should also be sensitive to
map weak magnetic fields in nanoscale.

\begin{figure}[h!]
\includegraphics[width=0.45\textwidth]{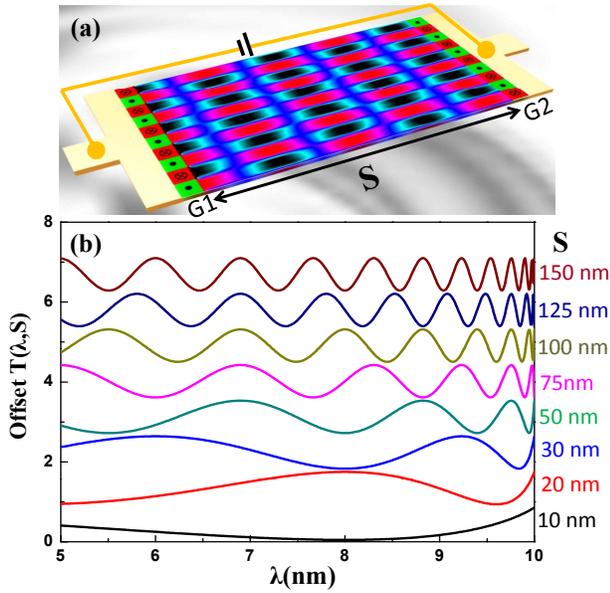}
\caption{(a) Schematic of quantum spin Talbot transistor. (b)
Characteristic quantum spin Talbot resistance depending on grating
separation $S$ and $\lambda$, $a$=10 nm and
0.5$<$$\zeta$$\leqslant1$.} \label{rys3}
\end{figure}

With electrodes connected to two gratings (G1 and G2), a quantum
spin Talbot transistor (QSTT) can be fabricated. By calculating
the spin transmission probability $T_{\pm}$$(\lambda,S)$ through
the second grating from Eq. \ref{probability_density} as an
approximation, we obtain \cite{wxt}
\begin{eqnarray}\label{resistor}
T(\lambda,S)=T_{+}+T_{-}=\frac{1}{4}+\frac{2}{\pi^{2}}+\frac{4}{\pi^{2}}\cos\frac{2\pi
S}{\mathcal{Z}_{T}(\lambda)}
\end{eqnarray}
\noindent where $S$ is the separation between G1 and G2. The
$T$($\lambda$,$S$) varies from single to multi-peak tuning via $S$
and $\lambda$ in the range 0.5$<$$\zeta$$\leqslant1$, exhibiting
behavior quite distinct in comparison to both GMR \cite{albert}
and spin Hall effect transistor \cite{spinhall}. By further
including the effect of electric field, we solve the
Schr$\ddot{o}$dinger equation with electrical potential, modifying
Eq.\ref{resistor} to include an Airy function $Ai(S)$ \cite{airy};
however, no dramatic change of the QSTT curve shape is expected
\cite{wxt}.

In conclusion, we propose an electron quantum spin Talbot effect.
Potential applications such as the QSTI and QSTT are present. The
success of the experiment will provide a new route to actualize
periodic spin state distributions in two-dimensional space, also
leading to spintronics applications which will be important for
future spin-based technologies.

We thank the referees for their insightful comments. WXT thanks
Dr. Zheng Gai at ORNL for illuminating discussions on potential
experimental realizations, and encouragement.


\begin{thebibliography}{99}
\bibitem{movie1}
In movie 1, the wavelength dependent QSTE is presented.
\end{thebibliography}
\end{document}